
\magnification=1200
\hoffset=-.1in
\voffset=-.2in
\vsize=7.5in
\hsize=5.6in
\tolerance 10000

\baselineskip 12pt plus 1pt minus 1pt
\def\thru#1{#1\!\!\!\!/}
\def\footnoterule{\kern -3pt \hrule width \hsize \kern 2.6pt}
\footline = {\ifnum\pageno>0 \hfil \folio \hfil \else\hfil\fi}
\quad

\quad

\vskip 48pt
\centerline{\bf TOWARDS A
SEMICLASSICAL SEISMOLOGY OF BLACK HOLES$^{*}$}\footnote{}{This
work is supported in part by funds
provided by the U.S. Department of Energy (D.O.E.) under contract
\#DE-AC02-76ER03069, and by the Division of Applied Mathematics of the U.S.
Department of Energy under contract \#DE-FG02-88ER25066.}
\vskip 48pt
\centerline{Michael Crescimanno}
\vskip 12pt
\centerline{\it Center for Theoretical Physics}
\centerline{\it Laboratory for Nuclear Science}
\centerline{\it and Department of Physics}
\centerline{\it Massachusetts Institute of Technology}
\centerline{\it Cambridge, Massachusetts\ \ 02139\ \ \ U.S.A.}
\vskip 1.5in
\centerline{Submitted to: {\it Phys. Rev. D\/}}
\vfill
\vskip -12pt
\line{CTP\#2226 \hfil hep-th/9312136 \hfil July 1993}

\pageno=0

\eject
\pageno=1

\baselineskip 24pt

\centerline{\bf ABSTRACT}
\medskip

Black hole spacetimes are
semiclassically not static.
For black holes whose lifetime is larger than the
age of the universe we compute, in leading order,
the power spectrum of deviations of the
electromagnetic charge from it's average value, zero.
Semiclassically
the metric itself has a
statistical interpretation and we
compute a lowerbound on its
variance.

\vfill
\eject

\noindent{\bf I.\quad INTRODUCTION}
\medskip
\nobreak

Since the seventies after the remarkable work of Hawking$^1$, Fulling$^2$
and others, there has been a flowering of technical development and interest
in semiclassical gravity\footnote{*}{By semiclassical gravity we mean
treating matter fields quantum mechanically in classical geometrical
backgrounds.}. The central problems of this approach, already outlined
in some of the very earliest papers, include understanding the meaning
of static solutions, the apparent loss of unitary evolution
in black hole spacetime,
and the nature of the endpoint of black hole evaporation.
It is difficult to ascertain whether these questions
are addressable with semiclassical techniques, and many
researchers have
found it instructive to
study these questions
in low-dimensional models such as two-dimensional dilaton gravity$^{3}$.

One of the central results of semiclassical gravity as it applies to
Schwarzschild spacetimes is that the Green function of quantum fields
is periodic in imaginary time. This means that one should regard the
fields outside the hole as being excited and
being characterized approximately in terms of a
thermal distribution.
An observer at rest in these co-ordinates but out in the asymptotically
flat region will see a thermal flux of particles emerging from the hole.
It is understood that this Hawking radiation is
responsible for the decay of the black hole.

Of course, since the black hole has a finite total energy, it's dissolution
into Hawking radiation is not a smooth process; in a fixed time interval the
amount of energy it radiates varies statistically.
The semiclassical result represents the time averaged rate of the
process.
Simply put, the Hawking
process is ''noisy''. Unfortunately,
the evaporation process is, by nature, a
non-equilibrium process and  would be
very difficult even in principle to model exactly.
In order to avoid non-equilibrium questions
we rigorously answer
what we feel are physically
interesting questions
about equilibrium states entirely within the context of
semiclassical gravity.
Since for
most of its evolution
the black hole spacetime is well
approximated
as a quasi-equilibrium state, assuming
equilibrium seems physically reasonable.
Indeed, the static classical behavior of
four-dimensional black holes in equilibrium with a
radiation bath is a well studied problem$^4$.
Furthermore, experience with
many dissipative systems even far from equilibrium
indicates that often the actual
noise has nearly the same norm and statistical properties
as noise computed assuming equilibrium conditions.

This work describes a computation of the power spectrum of the charge
of a four-dimensional black hole assuming equilibrium.
We do this in a regime where
the semiclassical response to charging the hole is given by a linear equation,
although this analysis may be extended to the non-linear regime.
The results we obtain
yield a lower limit on the the statistical variations
of the metric itself, which is interpretable as
the ''seismology'' of the spacetime's geometry.
We treat the metric as a background field and so
seismology arises in a different way than that of Ref.[5] where
the gravitational field is treated dynamically and semiclassically.
Section II contains an introduction to equilibrium
notions and noise, Section III is a computation of the power spectrum of
the charge for a black hole and Section IV discusses the physical
effects and back reaction on the metric.

\bigskip
\noindent{\bf II. \quad CHARGE AND EQUILIBRIUM}
\medskip
\nobreak

In this paper we will restrict our attention to nearly
static solutions of Einstein's equations coupled semiclassically
to the usual electromagnetic field and electron. Thus we
are interested in nearly static solutions that extremize
the action
$$ S = \int {d^4x} \sqrt{g} \big[ R + {{1}\over{4}} F^2 + i {\bar \psi}
({\thru{D}}-m)\psi\big].    \eqno(2.1)$$
We will show below that the results will not depend
on the particular type of charged matter
field that one uses, and only on the total number, $N$, of ''electron''
(lowest mass, charged) species.
For the purpose of simplicity here it
we require that all the lowest mass charged species have the
same charge.
Below, when discussing
numbers we use the physical (spin) degeneracy $N=2$ but
it is useful for reference to retain the $N$-dependence
explicitly in some of the formulae.

The static electrical properties of black holes are
summarized by the well known Reisner-Nordstrom (RN) solution to Eq.(2.1),
whose metric and gauge field are
(in the usual spherical co-ordinates $t,r,\theta,\phi$,)
$$ g_{\mu\nu} = diag\ ( B, -1/B, -r^2, -r^2 sin^2\theta),  $$
$$ A_\mu = (-Q/r,0,0,0)     \eqno(2.2)$$
where $B=1-2M/r+Q^2/r^2$, and with $M$ and $Q$ being the mass and charge
of the hole, respectively. The metric has two horizons,
$r_{\pm}= M \pm \sqrt{M^2-Q^2}$, and semiclassically the
temperature of the outgoing radiation as seen from
a stationary Schwarzschild observer at infinity is
$1/\beta= {{r_{+}-r_{-}}\over{4\pi r_{+}^2}}$.

The semiclassical
evolution of the RN solution has been studied
extensively. One of the earliest references is due to
Gibbons$^6$ where he computed the charge loss rate, mostly
with an interest in understanding the super-radiative regime
where $mM<eQ$ with $m$ and $e$ being the mass and charge of the
electron, the lightest charged species. This work was followed up by
many other authors$^{7,8,9,10}$, which studied in more generality and precision
the evaporation of charged black holes into empty space.
These works all use the simplest result of the Hawking calculation, namely
expressions for the time-averaged rate of energy or charge loss,
and the conclusions of these earlier works are summarized
phenomenologically in terms of decay rates for the
$M$ and $Q$ parameters of Eq.(2.1).
Thus, from the point of view of this simple use of semiclassical
gravity, one finds that the charge always dissipates away,
eventually
leaving a neutral hole. The main point of this paper will be
to show in a quantitative way that this expectation is too naive.
Instead, if we regard even an initially neutral hole as a
quasi-equilibrium state, the charge of
that hole will fluctuate about zero.
Let $<\ >$ represent an average over the equilibrium ensemble.
Although charge
symmetry for the spacetime of a neutral black hole suggests that $<Q> = 0$,
the charge on the hole will undergo statistical fluctuations so that
$<Q^2> \ne 0$. This fact was known long ago$^{8,11}$,
and may be understood as a
simple consequence of the equipartition theorem. Since the energy in the
electrostatic field outside of
a hole of mass $M$ and charge $Q<<M$
is
$$  U = {{Q^2}\over{2r_{+}}} \eqno(2.3)$$
we expect in equilibrium that this energy will be $\sim {{1}\over{2\beta}}$
so that
$$<Q^2> \sim  {{1}\over{4\pi\alpha}}  \eqno(2.4)$$
in electron units. Note that this result
is independent of the mass of the black hole,  and also
independent of the mass of the electron and the number of electron species.

It is important to emphasize that this $<Q^2>$ is not a consequence of
the granularity of charge, and is to be understood as a
statement about an ensemble of identically prepared
black holes (or equivalently, one black hole
observed over a long time) in equilibrium
with a heat bath. Thus, it is no difficulty that
$\sqrt{<Q^2>}\ne {\bf Z}$
in electron units. This result also holds for astrophysical
black holes in the limit of low total charge.
Of course, in most
practical applications, an astrophysical black hole in a space
plasma will tend to acquire a net positive charge
(which would be, in practice, much bigger than that
implied by Eq.(2.4))
as a consequence of the
well known Langmuir effect$^{12}$, which has nothing to do with
the effects of semiclassical gravity. We restrict our
study to that of a black hole in either empty space or a box filled
with thermal radiation
comprised solely of photons, electrons and positrons
but in thermal equilibrium with the hole.

\goodbreak
\bigskip
\noindent{\bf III \quad EQUILIBRIUM CORRELATION FUNCTIONS}
\medskip

The results of the previous section indicate that
the average charge of a nearly neutral black hole at equilibrium
is zero ($<Q>=0$) but that the average charge-squared $<Q^2>$ is
non-zero. If one throws charge into a
neutral black hole of
macroscopic horizon radius it will
stay charged for a very long time, and so
the above result, although correct, needs further elucidation.
In particular, it would be of most interest to calculate
the charge-charge correlation function $<Q(t)Q(0)>$. In this section we
compute this, and indeed semiclassically
all correlation functions of the
charge of a black hole in equilibrium
but only in the limit that
the total charge on the hole is small. By low charge we mean
the ratio, $\Lambda$, of electromagnetic force and gravitational force
on an electron is less than one; $\Lambda = eQ/mM < 1$.
Note that since $e/m >> 1$ in rationalized units, the
Hawking temperature for the RN spacetime (which is given above in terms
of $M$ and $Q$) is essentially independent of the charge.

In  computing this correlation function we will re-derive
and use a generalization of the
fluctuation-dissipation theorem (FDT)$^{13}$
for curved space. This theorem will ultimately
relate the charge-charge correlator ('fluctuation') to
a 'linear transport coefficient'  that
characterizes the evaporation
of the black hole's charge.
The linear transport coefficient for charge loss follows directly
from the results of Hawking, Fulling, Gibbons and others using the
quasi-equilibrium notions of semiclassical gravity.
It has long been known that the
best description of a black hole spacetime
semiclassically in equilibrium is as
an ensemble of spacetimes 'nearby' a given solution$^{11}$. Also, the
variance of this ensemble was computed by various means$^{11,14,15}$.
Page$^{8}$ computed the
likelyhood of any particular measurement of the charge of a
black hole, essentially finding
a gaussian probability with variance given by Eq.(2.4).

One way to understand the results
we
present here is that
we will compute more than just these 'static' (i.e. equal-time)
variances
such as Eq.(2.4). Instead of computing averages over the
ensemble (simple phase averages) we use a version of the FDT to
compute properties about the average {\it temporal} evolution
of a single system in the ensemble. Noise (that is, correlation
functions) in a co-ordinate of
a sub-system can be thought of as an
average over an ensemble of {\it trajectories} in the ensemble
used to compute phase averages. The temporal evolution
of the sub-system is a consequence
of the fact that it is dissipatively coupled to the
rest of the system.

It is very natural to understand the Hawking effect
as a dissipative phenomena. Dissipation is
generic to situation in which one studies
a particular degree of freedom (for example,
the metric or the charge of the hole)
of a system that is coupled to a continuum of other degrees
of freedom (here quantum fields)
whose precise evolution is treated approximately.
This connection between the Hawking effect and dissipation
has been explored mostly in the context of
isotropic cosmological models$^{16,17}$.

Strictly speaking, for
the FDT we develop and use here, we are formally
assuming that
the charge is a continuous variable and so are implicitly
working in the
approximation that $\alpha$ is small.
Versions of the FDT has been used before
in semiclassical general relativity
for the back-reaction on the gravitational shear due to
the emission of gravity waves$^{18}$ and studying the
stability of AdS spacetimes$^{19}$
but we use it in a rather different way here.

To begin with, consider that the number of charged quanta
of spin $s$ (in the
charged matter field of Eq.(2.1), $s=1/2$) emitted by
an isolated RN black hole (Eq.(2.2)) per unit time per unit
volume of final state phase space as seen by an asymptotic
Schwarzschild
observer is
$$ {{dn^{\pm}}\over{dt}} = {{N \sigma^{\pm}_E}
\over{e^{\beta(E\pm e\Phi_h)}
+ (-1)^{2s+1}} },
\eqno(3.1)$$
where $e$ is the charge of the matter field.
$\sigma^{\pm}_E$ is simply
the total absorption cross-section. The energy, $E$, is
$\sqrt{{\bf p}^2 + m^2}$
and $\Phi_h = Q/r_+$ is the 'voltage' of the horizon
w.r.t. infinity.
Gibbons$^6$ computed the rate of loss of charge from a black hole
by estimating $\sigma^{\pm}_E$ approximately via the
wave equation in the RN background. His work was later extended and
made more precise by
Ternov, et. al.$^7$. In the $\Lambda<1$ regime it is
especially easy to obtain their results with a classical computation
of $\sigma^{\pm}_E$ and since this will suffice in what follows
we briefly describe it's derivation.
Approximately, the total absorption cross-section $\sigma^{\pm}_E$ is
$$ \sigma^{\pm}_E =  {{A^{\pm}}\over{E^2-m^2}}   \eqno(3.2)$$
where $A^{\pm}$ is found by solving the
geodesic equation and to lowest non-vanishing
order in $M$ and $Q$ is
$$ A^{\pm} = 4\pi M^2 m^2(1 \mp \Lambda)+...         \eqno(3.3)$$
Now, using the fact that the total current at infinity
as given in terms of $dn^{\pm}/dt$ is
$$ {\dot Q} =e \int \big({{dn^{+}}\over{dt}} - {{dn^{-}}\over{dt}} \big)
{d^3{\bf p}} \eqno(3.4)$$
and furthermore, for holes whose lifetimes are
long compared with the present age of the universe and
using $m>1/\beta$
$$ {{1}\over{e^{\beta( E\pm e\Phi_h)} \pm 1}} \sim e^{-\beta (E\pm e\Phi_h)}
\eqno(3.5)$$
and working in the $\Lambda<1$ regime
we can use Eq.(3.2-3.5) to relate the net current at infinity
to the hole charge
$$  {\dot Q} =- Q /\tau   \eqno(3.6)$$
where
$$ \tau = {{\lambda^2}\over{8\pi\alpha c r_h N}}
e^{4\pi{{r_h}\over{\lambda}}}
\eqno(3.7)$$
is the lifetime of charge on the hole in terms of
$\lambda$, the compton wavelength of the
lightest charged particle, and $r_h$  ($\sim 2M$), the
Schwarzschild radius of the hole ($c$ is the speed of light.)
Notice that the charge loss is proportional to the total charge
and $\tau$ is only a function of $m$, $M$ ,the Planck scale and
$\alpha$. Note that this decay time
$\tau$ of Eq.(3.7)
is of a very different from that found in the $\Lambda >>1$ limit
(compare with Ref.[6,7,9].) This is due to the fact
that in the $\Lambda<1$ regime we ignore the small contribution
to the rate formula due to ''super-radiance'' that arises
from pair creation in the constant electromagnetic field$^{19}$.

To an observer in Schwarzschild co-ordinates, the evaporation of
the charge of the hole has a very simple
local explanation. Such an
observer would conclude
that since the quasi-equilibrium state
corresponds to putting the black hole in a box filled with
radiation at the temperature of the hole, the loss of charge
from the hole is simply a result of the charges in the bath
neutralizing the hole. Indeed, $\tau$ has the dimensions
of electrical resistivity and
it is precisely the resistance to the flow of the
electron and positrons in the geometrical background
of the hole.
It is instructive to
make this interpretation
quantitative. Note that the resistivity, $\rho$,
(supposing the metric is approximately a flat
Minkowski metric)
of a thermal bath is
given by
$1/\rho \sim n_e e {{\partial V}\over{\partial{\bf E}}}$
where $n_e$ is the number density of electrons and
positrons in the bath
$$ n_e \sim N\big({{ m}\over{\beta}}\big)^{3/2}
exp(-\beta m)  \eqno(3.8)$$
and ${{\partial V}\over{\partial{\bf E}}}$ is the drift
velocity per unit applied electric field. This may be easily
estimated by ${{\partial V}\over{\partial {\bf E}}} = el/mV$
where $V$ is the mean velocity of the electrons in the
bath ($\sim \beta ^{-1/2}$)
and $l$ is a length scale for scattering.
In the low temperature limit the scattering
due to the interaction of the drifting charges with the other
particles in the bath is negligible, and
$l$ is well estimated by a gravitational scattering length.
The gravitational scattering length is the length over
which a thermal particle
in the vicinity of the hole would experience an ${\cal O}(1)$
change in its velocity, $l \sim G_N M m \beta \sim M^2 m$
Putting this all together, we conclude
$$ \rho \sim {{e^{\beta m}}\over{\alpha m^2 M N}}   \eqno(3.9)$$
which has the same functional dependence as the
analytical result Eq.(3.7).

We now outline the general argument due to
Callen and Welton$^{13}$ as used by
Kubo$^{21}$
and generalized by
Martin and Schwinger$^{22}$ that relates the dissipation
described in Eq.(3.7) to the equilibrium
fluctuations in the charge-charge correlation
function.
On general grounds in the interaction picture
we may write the response (current)
to the presence of a
gauge field $A^\nu$ to first order in terms of the
commutator of two currents,
$$ <J_a> = <J_a>_0+\int <[J_a,J_\nu]> A^\nu +... \eqno(3.10)$$
where $<J_a>_0$ is the current when there is no applied
electric field (by charge invariance of our neutral black hole
state we ascertain that $<J_a>_0$ is zero) and where
roman indices refer only to space components (well defined
by the existence of a global time-like killing vector in the
spacetime of Eq.(2.2)) whereas
greek indices are Minkowski.

It is
convenient to define the local conductivity tensor
$\sigma_{a}^{b}(\omega)$ that relates the current $<J_a>$
to the total electric field $E_b =\int E_b(\omega)e^{-i\omega t} d\omega$
$$ <J_a(\omega)> = \sigma_a^b(\omega) E_b (\omega)  \eqno(3.11)$$
and by standard manipulations of Eq.(3.10), computing
as a local Schwarzschild observer would, we find
the Kubo$^{21}$ result,
$$ \sigma_a^b(\omega) = {{1}\over{\omega}}\int_0^{\infty}
dt e^{i\omega t} < \big[ J_a(t), J^b(0) \big] > + i{{\Gamma}\over{\omega}}
\delta_a^b
\eqno(3.12)$$
with $J_a$ being the current averaged
over space and
where $\Gamma$ is an infinite constant associated with the
spectral density of the photon propagator.
The term involving $\Gamma$ is a
trivial consequence of the
fact that, by gauge invariance, the equal time commutator of a
time and a space component of the electromagnetic current
cannot be zero. It is essentially the polarizability of the
vacuum modes, and of course cannot lead to any
dissipative effects$^{23}$. Now, although we might guess that
the commutator of two space components of the current is also
zero, in view of Eq.(3.6)
this can clearly not be the case in the curved space of Eq.(2.2).
We are interested in the part of the conductivity
$\sigma_a^b$ that is even in frequency $\omega$. By trivial
manipulations of Eq.(3.12) we have
$$ \sigma_a^b(\omega)+\sigma_a^b(-\omega) = {{1}\over{\omega}}
<\big[J_a,J^b\big]>(\omega)  \eqno(3.13)$$
As shown in Ref.[24], the correlation functions of
fields in a black hole spacetime behave
asymptotically
as
finite temperature Green functions at the temperature of the hole.
Thus, they satisfy the well-known KMS condition
and use of that condition allows one to relate
the fourier components of the commutator and
product
of two currents,
$$ <[J_a,J^b]> = (exp(\beta\omega)-1)<J_aJ^b>   \eqno(3.14)$$
The FDT in this context is just Eq.(3.13), the dissipation, taken
together with Eq.(3.14), $<J_a J^b>$ representing the fluctuation\footnote{*}
{As described later, for $\Lambda<1$ we will be most interested in the
high-temperature limit of the FDT and so the Wightman function
(RHS of Eq.(3.14)) will well approximate the causal Green function
that one would detect experimentally.}.
Thus integrating the
radial current in Eq.(3.6) that arises as a response to the
electric field (caused by the charge on the hole)
over a two sphere and putting all the metric factors
in Eq.(3.13), the
high temperature limit ($\beta \rightarrow 0$)
of Eq.(3.14) becomes,
$$ <Q_\omega Q_{-\omega}> \sim {{2}\over{\beta R\omega^2}}  \eqno(3.15)$$
where $R = \tau/r_+$ has dimensions of resistance.
Notice that this result diverges as $\omega \rightarrow 0$. Since
Parsifal's identity,
$$ <Q^2> ={{1}\over{2\pi}} \int <Q_\omega Q_{-\omega}> d\omega
\eqno(3.16)$$
for Eq.(3.15) implies that $<Q^2>$ is infinite which
one knows (see Eq.(2.4)) cannot be correct.
Indeed, in the exposition above
we have neglected the back reaction of the charge loss on the subsequent
evolution of the hole's charge. This may be seen most directly by
looking at the correlator as a function of time. Fourier transforming
Eq.(3.15) we find the correlation function
for brownian motion,
$$ <Q(t)Q(0)> = |t|/\beta R      \eqno(3.17)$$
This would imply that if you started with a neutral hole that
after some time the charge would have drifted quite far from
zero. This is what we would expect perhaps for the
evolution of a {\it global} charge of the hole but
not for a charge associated with a gauged symmetry.
Indeed, this Brownian-like behavior of the charge tells us that
we have unwittingly neglected the biasing
of the charge loss by the
electrostatic energy of the charge of the hole!

Thus far we have neglected the $F^2$ term in the Lagrangian
Eq.(2.1), and it is easy to see that
including it in the analysis will modify the
small $\omega$ behavior of Eq.(3.15).
The correct formula for the charge-charge
correlators may be
found by exploiting an
electrical analogy. From Eq.(2.3) above and from the
charge decay formula Eq.(3.6), it is natural to think of
a nearly neutral black hole as a circuit comprised of a
capacitor (of value $r_+$, the hole 'radius') shunted by a resistor
(of resistance $R = \tau/r_+$).
Here the resistor is
materially caused by the quantum fields outside the hole and
thus is in a quasi-equilibrium state at temperature $1/\beta$.
It is a well know consequence of the FDT
that a resistor at a high temperature/low frequency
acts as a source of 'white' voltage noise,
$$ <V_\omega V_{-\omega}> = R/\beta     \eqno(3.18)$$
and so the equivalent
circuit diagram of our nearly neutral black hole is as shown in
figure 1. Using Kirchoff's laws it is now simple to ascertain the
charge-charge correlator
$$ <Q_\omega Q_{-\omega}> = {{R r_+^2}\over{\beta(1+\omega^2\tau^2)}} \ .
\eqno(3.19)$$
This formula is finite as $\omega\rightarrow 0$ and as expected
yields Eq.(2.4) when integrated via Eq.(3.16)\footnote{*}{Actually, Eq.(3.19)
is the high temperature limit of the full FDT result
$$<Q_\omega Q_{-\omega}>= {{R r_+^2 \omega}\over{(1+\omega^2\tau^2)
(\exp(\beta\omega)-1)}}.$$
Most of the support for this correlation function comes from the
region $\omega<1/\tau$ and since for $\Lambda<1$, $\beta/\tau\sim 0$,
Eq.(3.19) is a very good approximation to the full FDT result.}.
Also, for a linear system, such as the loss of charge off
of a nearly neutral hole for $\Lambda < 1$, all
higher correlation functions of the charge descend from
the two-point correlator in the usual way. This result generalizes
the equal-time correlators of Page$^{8}$.

\goodbreak
\bigskip
\noindent{\bf IV. \quad CONSEQUENCES AND A SEISMIC LOWER BOUND}
\medskip

We now discuss a few consequences of the power spectrum
found in the previous section for the charge
of a nearly neutral black hole. We first discuss some issues
about the classical phenomenology of this noise spectrum and
finally discuss questions about its effect on the structure
of the spacetime itself.

The correlation function Eq.(3.19) depends on a single time scale
$\tau$. As seen from Eq.(3.7) this time scale can be truly huge for
astrophysical black holes. In those cases, all the 'noise' of Eq.(3.19)
happens at incredibly low frequencies and for all practical purposes
we would say that the charge of the black hole was a constant. However,
for a medium-sized hole whose lifetime is still long compared with the
lifetime of the universe the noise can
have support at 'everyday' frequencies.
For example, for a black hole that would have a lifetime of
a hundred times the
age of the observable universe we find that $\tau \sim 10^{-6}$ seconds
and by Eq.(3.19) that the hole charge is 'noisy' into megahertz frequency
range.

What would be a practical observable
consequences of this charge noise?
It is interesting to calculate the effect this
stochastic charge of the
black hole would have on the scattering of a slowly (with respect
to $r_+/\tau$) moving
particle of fixed charge. One finds that the noise
has a tendency to repel charges,
regardless of their sign, away
from the hole! This rather curious-sounding conclusion can be reached a
variety of ways, perhaps most instructively by
integrating over the Gaussian noise kernel Eq.(3.19) in the
path integral for the test particle.
The leading effect may be understood by
paying particular attention to the zero-temperature limit
(see the footnote below Eq.(3.19)),
as one does in the related case of dissipative quantum mechanics$^{25}$
when interested in the classical limit of the effective
equations of motion. The path integral for a test particle of unit charge
($\pm$)
moving under the influence of a stochastic charge $Q$ is,
$$ Z = \int[dQ] [d{\bf x}] {\rm M}(Q) exp\int dt{\bigg \{} {{m}\over{2}}
{\dot{\bf x}}^2\pm {{Q}\over{|{\bf x}|}}{\bigg \}}
\eqno(4.1)$$
Where we have for simplicity ignored the gravitational
attraction of the hole.
${\rm M}(Q)$ is the gaussian density whose variance is given
by Eq.(3.19), in the low temperature limit.
Performing the integration over $Q$ we find
$$ Z = \int[d{\bf x}] exp\int dt{\bigg \{}{{m}\over{2}}
{\dot{\bf x}}^2 +{{1}\over{2R}}
\int dt'{{1}\over{|{\bf x(t)}|}}
\theta(t-t'){{1}\over{|{\bf x(t')}|}}{\bigg \}}
\eqno(4.2)$$
Writing down the equations of motion that follow
from this lagrangian, we see that the last term contributes
a term that is odd under $t \rightarrow -t$.
Non-local terms ('memory') in equations of motion
are typical of the low-temperature limit of
dissipative systems (see for example Ref.[24].)
The net effect of the fluctuations of the charge of the
hole is to create an effective repulsive force for charged
test particles
of either sign!
As expected, the smaller the hole, the larger repulsion is for
slowly moving test charges. Furthermore,
the repulsion is proportional to the
square of the charge of the particle.
This behavior may be understood
qualitatively from very simple models and
is borne out well quantitatively in computer simulations
of the motions of test charges under the influence
of a stochastic charge of this type.

{}From the point of view of the correlation function Eq.(3.19) and the
charge loss equation Eq.(3.6) we may make a simple point about
the classical version of the information loss paradox. The classical version
of the information loss paradox
in this context asks if it is possible to
reproduce the initial time-dependent current that was used
to charge a hole given only the resulting evaporation current.
Indeed, Eq.(3.19) informs us that we should expect there to be
all sorts of temporal correlations in the evaporation current
that don't carry any 'information' about the initial charging
current but that are a result
of the quasi-equilibrium state.
Furthermore, the first-order equation for the
charge loss indicates that it will not be possible to simply
'evolve back' the fourier modes of the measured evaporation current.
Thus, as expected in this semiclassical picture, it is not
possible to recover the information about the time-dependence of the
initial charging current: electrically black holes act as
lossy low-pass filters. Of course, one hopes that understanding
evaporation non-perturbatively will ameliorate both this classical
and the quantum mechanical information loss problem.

Another interesting consequence of the stochastic nature of
the charge of the black hole is that the metric inherits
a statistical interpretation.
The metric of Eq.(2.2) for a static charged
black hole has $g_{00}=1-2M/r+Q^2/r^2$,
and since we expect $Q^2$ to vary
in accordance with Eq.(3.19)
we expect that as a consequence of back-reaction,
the metric itself
has statistical dispersion.
Rather than solve the Vadia-Bonner metric
for the time dependent metric exactly we approximate the true
solution by the RN solution with time dependent
$M$ and $Q$.
Indeed, for a small hole
in a box filled with thermal radiation,
thermal equilibrium suggests
$<\delta M>=0$ and we expect that
the noise in the mass fluctuations
has most of its support at lower frequencies\footnote{*}
{We would expect this to be true whenever the
timescale $\tau$ describing the charge loss is small
compared with the lifetime of hole were it to radiate
into an empty vacuum. This is precisely the case for a
black hole whose lifetime is shorter than about $10^4$ times
the present age of the universe.}
Thus, the total metrical dispersion, in the high frequency ($\omega>1/\tau$
but $\omega<1/\beta$, beyond which all noise is exponentially damped anyways.)
limit, should be strictly larger than that due to
the charge-charge correlator Eq.(3.19)
$$ g_{00}(\omega) = \int dt e^{i\omega t} g_{00}(t)$$
$$ <g_{00}(\omega)g_{00}(-\omega)>\ \  > {{\tau}
\over{4+\omega^2\tau^2}} {{1}\over{r_1^2 r_2^2}} \eqno(4.3)$$
Seismology is the study of the temporal evolution of the geometry
of quasi-static objects. Eq.(4.3) may be thought of as
a lower bound for the seismic activity in the geometry of
the spacetime of a black hole, as seen by a stationary
asymptotic observer. If one interprets Eq.(4.3) and the
correlator Eq.(3.19) as noise in the radial position of
causal horizon, we find that the horizon dilates
out and in
several multiples of the hole's compton wavelength.
These are very small 'quakes' but correspond to
radial dilation that could be larger than the distance between
the horizon and the stretched horizon
and so may be of consequence for understanding
what an asymptotic observer sees happening at the
stretched horizon$^{26,27}$.
Note that if we fix $\omega$
and pass to the thermodynamic limit ($M \rightarrow \infty$)
these fluctuation vanish, as expected on general
principles.
also, although Eq.(4.3) is second order in the temperature
(that is, second order in Planck's constant) it is the
leading term and going beyond one loop in the quantum fields
will generate higher order terms.
Also, these radial dilations
are completely distinct from the dispersion in the position
of $r=0$ due to momentum conservation
during evaporation of Hawking quanta.
They are also distinct from the dispersion of the gravitational
shear of Ref.[18]. It is
straightforward to compute the 'seismology' of the hole's
geometry in higher spherical harmonics, but that
is technically more complicated.

\goodbreak
\bigskip
\noindent{\bf V. \quad LANGEVIN VIEW OF SEMICLASSICAL GRAVITY}
\medskip

Of course not only the charge $Q$ but also, $M$, the mass of the hole
evolves semiclassically, and to really compute the
correlations of the metric
response we need to all the correlators
$<MM>$, $<MQ>$, $<QQ>$ (we neglect angular momentum
evolution to simplify the discussion) {\it and then}
compute the linear response about the metric
Eq.(2.2) using the Einstein equations.
This is a Langevin-like approach
to the back reaction in
Einstein equations. Usually, in studying the back reaction
semiclassically one solves
$$ G_{\mu\nu} = <T_{\mu\nu}>  \eqno(5.1)$$
where $<T_{\mu\nu}>$ is the renormalized stress tensor
in which appropriate boundary conditions are taken
to match with the appropriate physical situation, for example,
the evaporation of the hole. Since the spacetimes one
is most interested in
often have finite asymptotic masses and
charges, we have argued in this paper that it is important to
take into account the fluctuations about these values.
The fluctuations for a black hole radiating into empty
space are likely to be very close to those computed assuming equilibrium.
Thus, in this Langevin-like approach
to semiclassical gravity we would replace
Eq.(5.1) with Einstein's equation
$$ G_{\mu\nu} = T_{\mu\nu}  \eqno(5.2)$$
and to solve this for the variables
$g_{\mu\nu}$ and their statistical properties,
we specify
$<T_{\mu\nu}>$ and $<T_{\mu\nu}(x)T_{\alpha\beta}(x')>$
and all the corelation functions of the source
$T$ assuming equilibrium.
This is difficult to do in practice for technical
reasons.
In this paper we have not solved Eq.(5.2) but
the technically simpler situation of Einstein-Maxwell where we have
treated only the electromagnetic current and its back reaction on
the metric as a Langevin system. Understanding
semiclassical gravity from this quasi-equilibrium viewpoint
which philosophically treats back reaction as a
Langevin-like problem could lead to further elucidation of
the properties of black holes.

\goodbreak
\centerline{\bf Acknowledgements}
\medskip

The author has profitted greatly from conversations with
S. Axelrod, D. Freed, D. Freedman, G. W. Gibbons, R. Jackiw,
E. Keski-Vakurri, A. Landsberg, S. Mathur, M. Ortiz, D. N. Page, M. Perry
and I. M. Singer. The author wishes to thank J. W. York, Jr.
for pointing out the connections between this work and Ref.[5].

\vfill
\eject
\centerline{\bf REFERENCES}
\medskip
\medskip
\item{1} S. W. Hawking, {\it Commun. Math. Phys.\/} {\bf 43}, 199 (1975).
\medskip
\item{2} S. A. Fulling, {\it Phys. Rev.\/} {\bf D7}, 2850 (1973).
\medskip
\item{3} C. Callan, S. B. Giddings, J. A. Harvey
and A. Strominger, {\it Phys. Rev.\/} {\bf D45}, 1005 (1992),
R. B. Mann and T. G. Steele, {\it Class. Quant. Grav.\/}
{\bf 9}, 475 (1992);
for charged black hole solutions in 2-d see,
M. D. McGuigan, C. R. Nappi, and S. Yost, {\it Nucl. Phys.\/}
{\bf B375}, 421 (1992),
G. W. Gibbons and M. J. Perry, {\it Int. J. Mod. Phys.\/} {\bf D1}, 335 (1992),
C. R. Nappi and A. Pasquinucci, {\it Mod. Phys. Lett.\/} {\bf A7}, 3337 (1992).
\medskip
\item{4} G. W. Gibbons and M. J. Perry, {\it Proc. R. Soc. London\/}
{\bf A358}, 467 (1978),
D. Page, {\it Black Hole Physics}, V De Sabbate and Z. Zhang,
{\it eds.} 185 (1992), D. Hochberg, T. W. Kephart and J. W. York, Jr.,
Preprint IFP-467-UNC.
\medskip
\item{5} J. W. York, Jr., {\it Phys. Rev.\/} {\bf D28}, 2929 (1983).
\medskip
\item{6} G. W. Gibbons, {\it Commun. Math. Phys.\/} {\bf 44}, 245 (1975).
\medskip
\item{7} I. M. Ternov, A. B. Gaina and G. A. Chizov, {\it Sov. J.
Nucl Phys.\/} {\bf 44}, 343 (1986).
\medskip
\item{8} D. N. Page, {\it Phys. Rev.\/} {\bf D16}, 2402 (1977).
\medskip
\item{9} W. A. Hiscock and L. D. Weems, {\it Phys. Rev.\/}
{\bf D41}, 1142 (1990).
\medskip
\item{10} Y. Kaminaga, {\it Class. Quantum Grav.\/} {\bf 7}, 1135 (1990).
\medskip
\item{11} J. D. Beckenstein, {\it Phys. Rev. D.\/} {\bf 12}, 3077 (1975).
\medskip
\item{12} I. Langmuir and L. Tonks, {\it Phys. Rev.\/}
{\bf 34}, 876 (1929), W. B. Thompson,
{\it An Introduction to Plasma Physics}, (Addison-Wesley, Reading, MA)
1962.
\medskip
\item{13} H. B. Callen and T. A. Welton, {\it Phys. Rev.\/}
{\bf 83}, 34 (1951).
\medskip
\item{14} D. Pavon and J. Rubi, {\it Phys. Rev.} {\bf D37}, 2052 (1988).
\medskip
\item{15} D. Pavon, {\it Phys. Rev.\/} {\bf D43}, 2495 (1991).
\medskip
\item{16} M. Morikawa, {\it Phys. Rev. D\/} {\bf 33}, 3607 (1986).
\medskip
\item{17} E. Calzetta and B. L. Hu, {\it Phys. Rev.\/}
{\bf D40}, 656 (1989).
\medskip
\item{18} P. Candelas and D. W. Sciama, {\it Phys. Rev. Lett.\/} {\bf 38}
,1372 (1977).
\medskip
\item{19} E. Mottola, {\it Phys. Rev.\/} {\bf 33}, 2136 (1986), Preprint
LA-UR-92-1238.
\medskip
\item{20} J. S. Schwinger, {\it Phys. Rev.\/} {\bf 82}, 664 (1951).
\medskip
\item{21} R. Kubo, {\it J. Phys. Soc. Japan\/} {\bf 12}, 570 (1957),
G. D. Mahan, {\it Many Particle Physics},  (Plenum, New York) 1990.
\medskip
\item{22} P. C. Martin and J. Schwinger, {\it Phys. Rev.\/}{\bf 115},
1342 (1959).
\medskip
\item{23} C. Itzykson and J.-B. Zuber, {\it Quantum Field Theories},
(McGraw-Hill, New York) 1980; S. Trieman, R. Jackiw and D. Gross,
{\it Lectures on Current Algebras and Their Applications},(Princeton
University Press, Princeton, N.J.) 1972.
\medskip
\item{24} G. W. Gibbons and M. J. Perry, {\it Phys. Rev. Lett.\/} {\bf 36}
, 985 (1976).
\medskip
\item{25} A. O. Caldiera and A. J. Leggett, {\it Ann. Phys. (N.Y.)\/}
{\bf 149}, 374 (1983).
\medskip
\item{26} K. S. Thorne, R. H. Price and D. A. McDonald,
{\it Black Holes: The Membrane Paradigm}, (Yale University Press, New
Haven, Conn.) 1986.
\medskip
\item{27} L. Susskind, L. Thorlacius and J. Uglum, hepth/9306069.
\end